\documentstyle[times,pramana,epsf,floats]{ias}

\begin{document}
\mark{{Two Pion Production in Photon Induced Reactions}{S.Schadmand}}
\title{Two Pion Production in Photon Induced Reactions}
\author{S.Schadmand}
\address{Institut f\"ur Kernphysik, Forschungszentrum J\"ulich, Germany}


\abstract{
Differences in the photoproduction of mesons on the free proton
and on nuclei are expected to reveal changes in the properties of
hadrons.
Inclusive studies of nuclear photoabsorption
have provided evidence of medium modifications.
However, the results have not been explained
in a model independent way.
A deeper understanding of the situation is anticipated from a
detailed experimental study of meson photoproduction from nuclei
in exclusive reactions.
In the energy regime above the $\Delta$(1232) resonance,
the dominant double pion production channels are of particular interest.
Double pion photoproduction from nuclei is also used
to investigate the in-medium modification of meson-meson interactions.
}

\maketitle
\section{Introduction}

The study of in-medium properties of mesons and nucleon resonances
carries the promise to find signatures for partial chiral symmetry
restoration at finite baryon density and temperature.
Initially,
the scaling law proposed by Brown and Rho indicated a direct connection
between the vector meson masses and the chiral condensate
\cite{Brown:1991kk}.
This prospect has caused great interest in the properties of light mesons
in a dense and hot environment
\cite{Asakawa:1993pq,Rapp:1999ej,Post:2001am,Cabrera:2000dx}.
Various theoretical predictions indicate that the observation of
an in-medium  modification of the vector meson masses can provide
a unique measure for the degree of chiral symmetry breaking
in the strongly interacting medium \cite{Hatsuda:1992ez,Hatsuda:1993bv}.
However, in \cite{Leupold:1998bt}, it is shown that QCD sum rules
could rather be fulfilled by increasing the width of the hadron in medium.
In both scenarios, it is expected that hadronic strength is
shifted towards lower masses.
Some experimental observations are consistent with a modification of
the $\rho$ resonance in the nuclear medium \cite{Adamova:2002kf,Adams:2003cc}.
Recently, an indication for a downward mass shift of the $\omega$ meson
has been observed in photon-induced reactions on nuclei \cite{Trnka:2005ey}.

\section{Photoabsoprtion}

Photoabsorption experiments on the free nucleon  demonstrate the complex structure
of the nucleon and its excitation spectrum.
\begin{figure}[htb]
\epsfxsize=0.6\linewidth
\centerline{\epsfbox{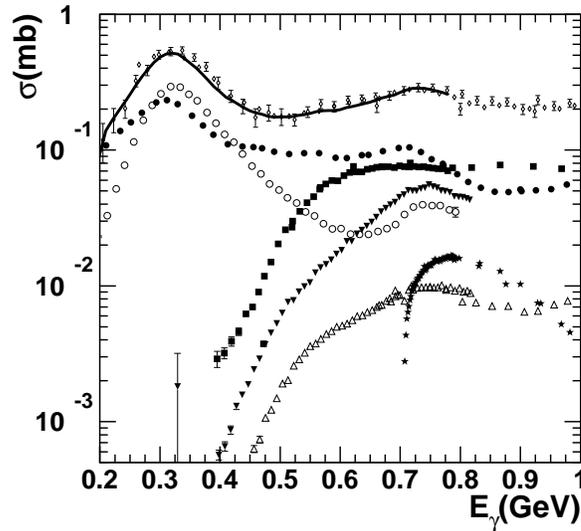}}
\caption{
 Photoabsorption cross section on the proton
 and decomposition into meson production channels.
 Small open circles are the photoabsorption data compilation 
 from \protect\cite{Hagiwara:2002fs}.
 The experimental meson production cross sections are:
 single $\pi^+$     (solid circles) from 
 \protect\cite{Buechler:1994jg,MacCormick:1996jz},
 single $\pi^\circ$ (open circles) from \protect\cite{Harter:1997jq},
 $\pi^+\pi^-$ (solid squares) from \protect\cite{Braghieri:1995rf,Wisskirchen-doc},
 $\pi^+\pi^\circ$  (downward solid triangles) from 
 \protect\cite{Langgartner:2001sg},
 $\pi^\circ\pi^\circ$ (upward open triangles) from 
 \protect\cite{Kotulla-doc,Hourany-inpc-2001}, and
 $\eta$ (stars) from \protect\cite{Krusche:1995nv,Renard:2000iv}.
 The solid line is the sum of the meson channels up to 800~MeV.
}\label{fig:photo-decomp-p}
\end{figure}
The lowest resonance is called $\Delta$(1232) which is a P$_{33}$ state
in the common notation (L$_{(2I)(2J)}$) with a pole mass of 1232~MeV.
It is prominently excited by incident photons of 0.2--0.5~GeV.
The following group of resonances, P$_{11}$(1440), D$_{13}$(1520), and S$_{11}$(1535),
is called the second resonance region (E$_\gamma$=0.5-0.9~GeV).
The observed resonance structures
have been studied using their decay via light mesons, showing that the photoabsorption
spectrum can be explained by the sum of $\pi$, $\pi\pi$ and
$\eta$ production cross sections.
Fig.~\ref{fig:photo-decomp-p}
shows the photoabsorption cross section
on the proton along with the experimental meson photoproduction cross sections.
The shapes of the meson cross sections reflect the resonance structures
observed in photoabsorption showing that the mesons are mostly decay products
of the respective resonances.
Single pion production is dominant in the region of the $\Delta$(1232) resonance.
Also, the three resonances
comprising the second resonance region, decay to $\sim$50\% via single 
pion emission.
This fact has been extensively exploited in partial wave analyses.
Above E$_\gamma\approx$0.4~GeV, the photoproduction of two pions is kinematically
possible and single $\pi$ production looses in dominance.
The solid line in Fig.~\ref{fig:photo-decomp-p} represents
the sum of the meson cross sections up to 0.8~GeV and demonstrates
that the photoabsorption cross section on the proton can be
explained by its decomposition into meson production.

The $\eta$ production threshold is located at E$_\gamma\approx$700~MeV.
The steeply rising $\eta$ cross section in Fig.~\ref{fig:photo-decomp-p}
is characteristic for an s-wave resonance.
The angular distributions of the $\eta$ emission are consistent
with this observation and
the cross section peaks around the mass pole of the S$_{11}$(1535)
resonance~\cite{Krusche:1995nv,Renard:2000iv}.
This resonance is unique in the sense that is has a strong decay branch
of 30-55\% into $\eta$ mesons.
Thus, $\eta$ production is considered characteristic for the S$_{11}$(1535) 
resonance.
Above the $\eta$ threshold, the cross section basically displays the resonance 
line shape enabling detailed studies of that state.
However, the contribution to the total is small.

\section{Nuclear Photoabsoprtion}

Photon induced reactions are particularly well suited to study
in-medium effects
in dense nuclear matter since photons probe the entire nuclear volume.
The first experimental investigation of the nuclear response to photons
was performed with total photoabsorption measurements from nuclei
with mass numbers ranging from $^7$Li to $^{238}$U.
The nuclear cross sections are practically identical
when scaled by the atomic mass number, thus scaling expectedly
with the nuclear volume.
However, the measurements
indicate a depletion of the resonance structure in the
second resonance region \cite{Frommhold:1992um,Frommhold:1994zz,Bianchi:1994ax}.
Bianchi et al. \cite{Bianchi:1994ax} reported that,
while in the $\Delta$-resonance region strength is only redistributed
by broadening effects, strength is missing in the D$_{13}$(1520)
region.
This observation has been taken has been one of the first
indications of a medium modification.

\begin{figure}[hbt]
\epsfxsize=0.5\linewidth
\centerline{\epsfbox{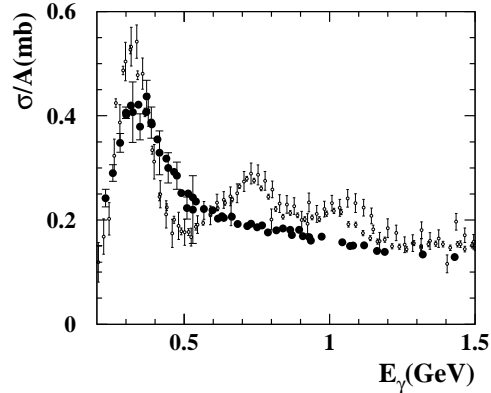}}
\caption{
Nuclear photoabsorption cross section per nucleon
as an average over the nuclear systematics~\protect\cite{Muccifora:1998ct}
(full symbols) compared to the absorption on the proton
\protect\cite{Hagiwara:2002fs} (open symbols).
}\label{fig:photoabs-nucs}
\end{figure}

Fig.~\ref{fig:photoabs-nucs} shows the nuclear photoabsorption
cross section per nucleon
as an average over the nuclear systematics~\cite{Muccifora:1998ct}.
The $\Delta$ resonance is broadened and slightly shifted while
the second and higher resonance regions seem to have disappeared.

Mosel et al.~\cite{Mosel:1998rh},
have argued that an in-medium broadening of the D$_{13}$(1520)
resonance is a likely cause of the suppressed photoabsorption cross
section.
Recent calculations are based on the BUU equation and are described in
\cite{Lehr:1999zr,Effenberger:1997rc,Muhlich:2004zj}.
Hirata et al.~\cite{Hirata:2001sw} have discussed a change
of the interference effects in the nuclear medium as one of
the most important reasons for the suppression of
the resonance structure.

It may be concluded that inclusive reactions like total photoabsorption
do not allow a detailed investigation of in-medium effects.
A deeper understanding of the situation is anticipated from the
experimental study of meson photoproduction on
nucleons embedded in nuclei in comparison to studies on the free nucleon.

\section{Double Pion Photoproduction}

Single as well as double pion production channels display
structure at the corresponding resonance mass,
i.e. around E$_\gamma\approx$760~MeV.
As could already be seen from Fig.~\ref{fig:photo-decomp-p},
the meson production channels involving charged pions are dominant
as expected in electromagnetic excitation processes.
Two-pion production is characteristic for the D$_{13}$(1520) and
P$_{11}$(1440) resonances.
Furthermore, the dominant resonance contribution comes from
the D$_{13}$(1520) resonance
having the strongest coupling to the incident photon.
Because of the importance for reactions on the proton,
it is expected that double pion production plays an important role in
the understanding of the medium modifications as observed in nuclear
photoabsorption.

\begin{figure}[htb]
\centerline{
 \epsfxsize=0.5\linewidth \epsfbox{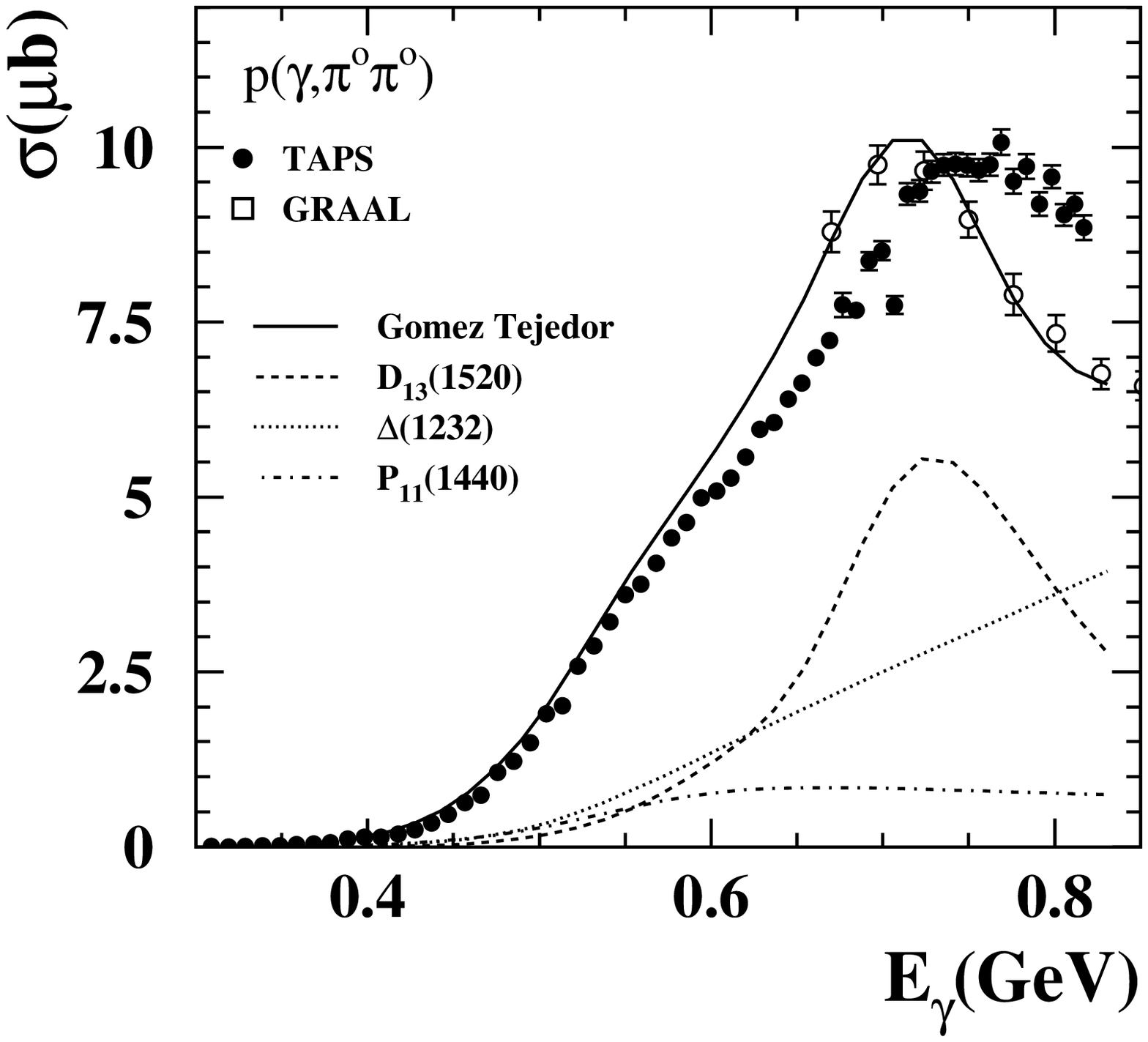}
 \epsfxsize=0.5\linewidth \epsfbox{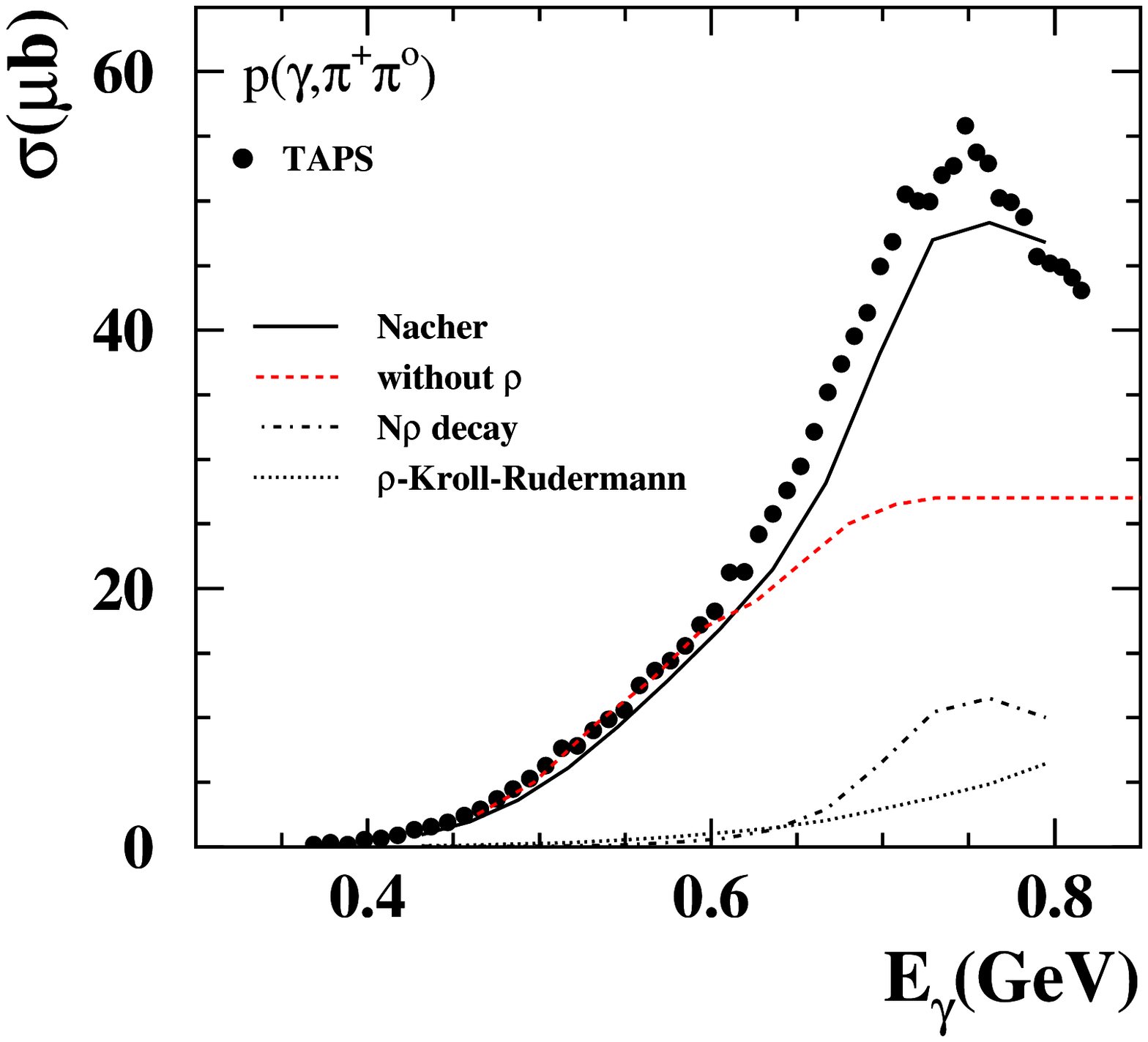}
}
 \caption{
 $\pi^\circ\pi^\circ$ (left) and $\pi^\circ\pi^+/-$ (right)
 photoproduction from the proton \protect\cite{Krusche:2003ik}.
 Calculations by \protect\cite{Nacher:2000eq}.
 }\label{fig:pipi-p}
\end{figure}
On the proton, three isospin combinations of pion pairs
can be produced.
The corresponding cross sections are shown in Fig.~\ref{fig:pipi-p}
along with theoretical calculations \cite{Nacher:2000eq}.
The study of  $p(\gamma,\pi^\circ\pi^\circ)$ (left panel
of Fig.~\ref{fig:pipi-p}) revealed
that $\Delta$ intermediate states are important.
The $N^*$ contribution to double pion photoproduction
by itself is not large but rather stems from an interference with
other terms \cite{GomezTejedor:1996pe,Nacher:2000eq}.
A similar behavior is found in $(\gamma,\pi^+\pi^\circ)$ reactions,
shown in the right panel of Fig.~\ref{fig:pipi-p}.
Additionally, the peak in the $(\gamma,\pi^+\pi^\circ)$
cross section can only be
explained by contributions from $\rho$ production terms,
with a decay branch of 20\% for D$_{13}\to N\rho$ \cite{Langgartner:2001sg}.
This decay mode is forbidden in $(\gamma,\pi^\circ\pi^\circ)$ reactions.

\section{Double Pion Photoproduction from Nuclei}

\begin{figure}[htb]
  \centerline{
   \epsfxsize=0.5\linewidth \epsfbox{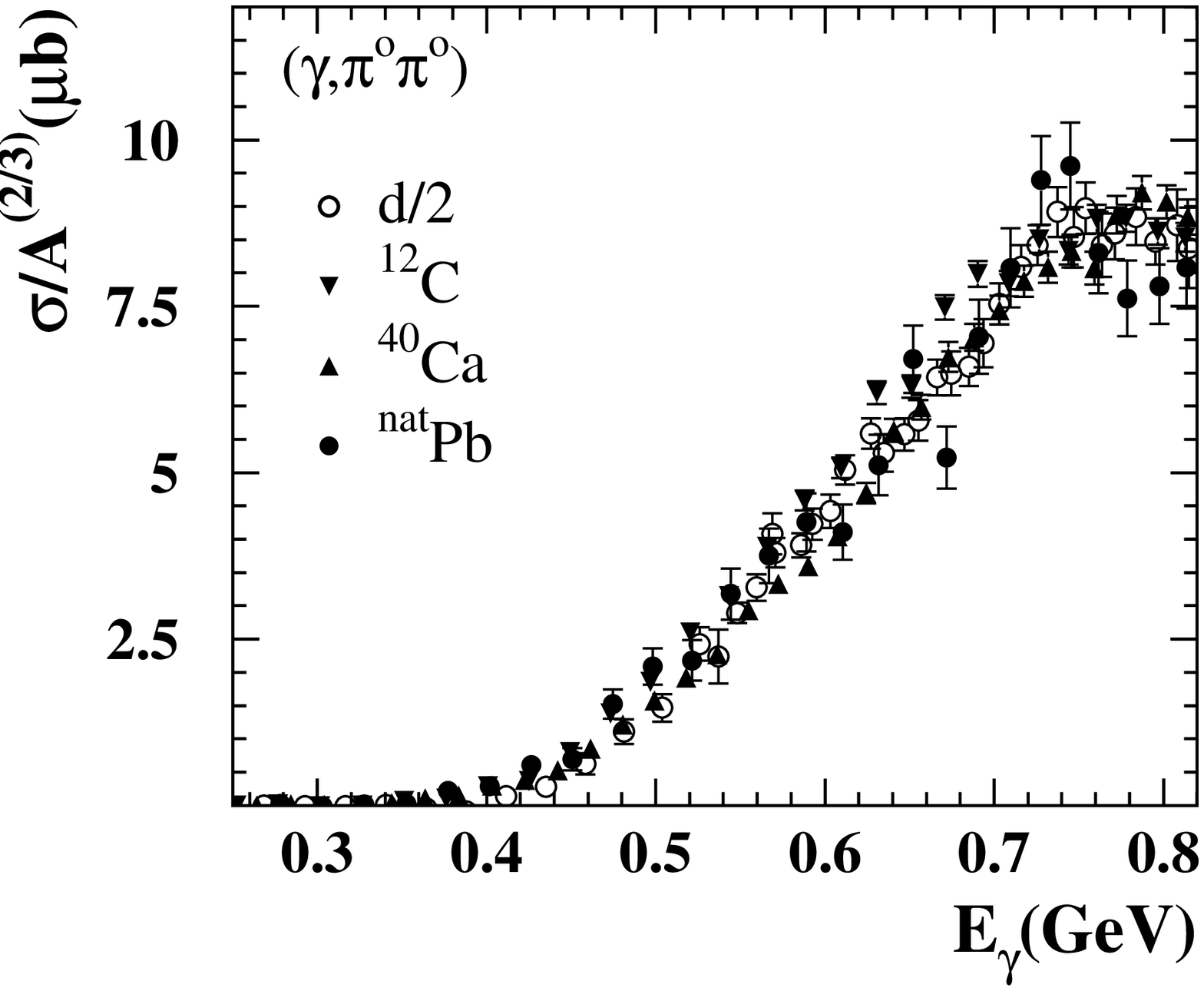}
     \epsfxsize=0.6\linewidth \epsfbox{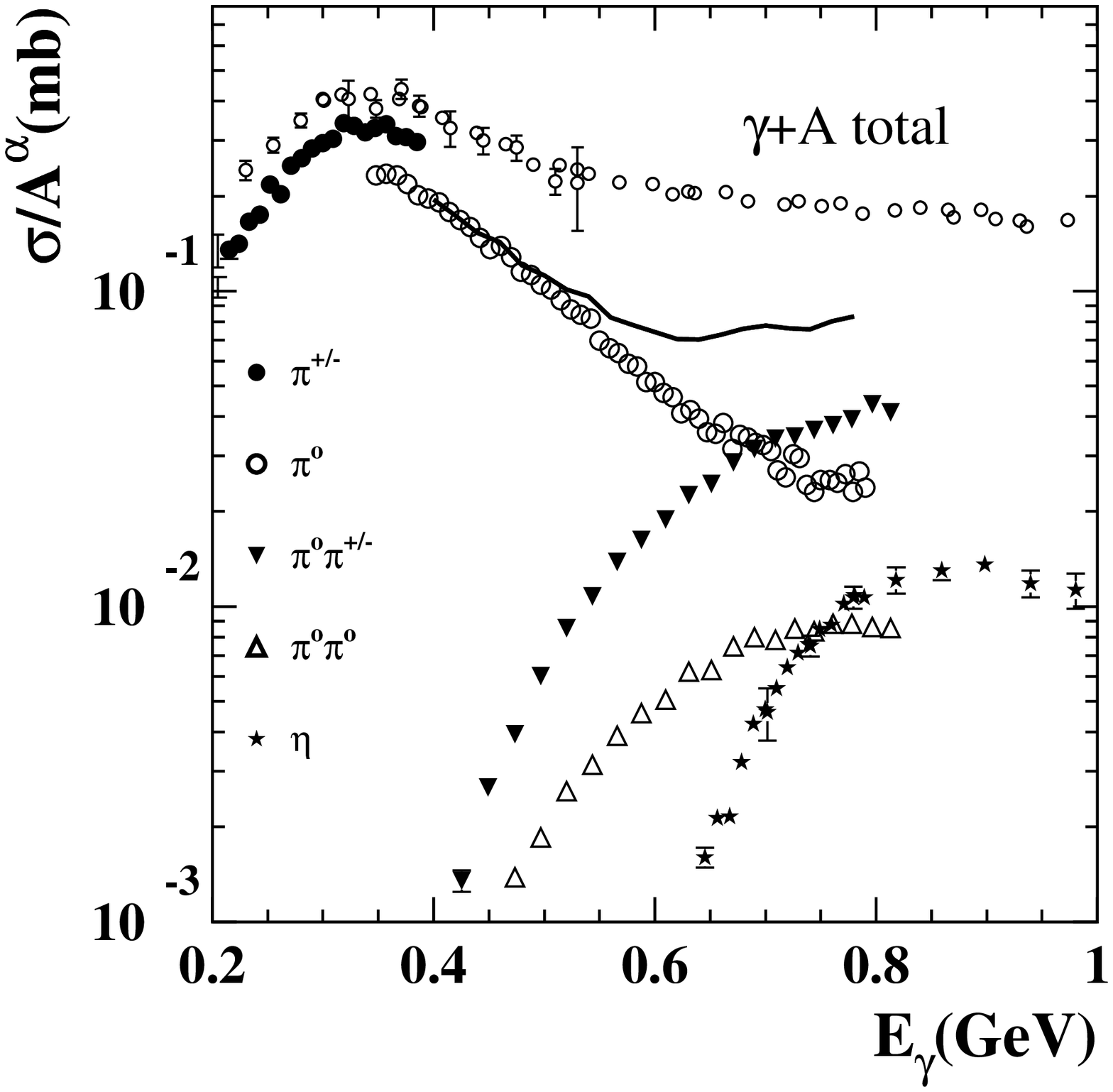}
   }
 \caption{
  Left:
  Preliminary total cross sections for $\pi\pi$
    photoproduction from different nuclei \protect\cite{Krusche:2004zc} along with
    results from the deuteron~\protect\cite{Kleber:2000qs}.
    The nuclear cross sections are divided by A$^{2/3}$,
    the $\pi^\circ\pi^\circ$ deuteron cross section by 2.  
  Right: 
  Status of the decomposition of nuclear photoabsorption into
  meson production channels (scaled with A$^\alpha$, $\alpha$=2/3.
  Small open circles are the average nuclear photoabsorption cross section
  per nucleon ($\alpha$=1) \protect\cite{Muccifora:1998ct}.
  Meson production data are from a carbon target
  \protect\cite{Krusche:2004zc,Arends:1982ed,Krusche:2001ku,Roebig-Landau:1996xa,Yamazaki:2000jz}. 
  The solid line is the sum of the available meson cross sections
  between 400 and 800~MeV.
 }\label{fig:1}
\end{figure}

The left panel of Fig.~\ref{fig:1} shows preliminary cross sections for
$\pi^\circ\pi^\circ$ photoproduction on a series of nuclei
from a recent analysis \cite{Janssen-02} of data taken with the TAPS detector 
at the MAMI-B accelerator (Mainz, Germany).
The nuclear cross sections are divided by A$^{2/3}$ and
compared to results from nucleons bound in deuterons.
With a scaling of A$^{2/3}$, the nuclear data agree almost exactly
with the cross sections on the nucleon.
Thus, the total nuclear $\pi\pi$ cross sections
do not seem to show any modification beyond absorption effects.

The right panel of Fig.~\ref{fig:1} summarizes
the systematic study of the total production cross sections for single
$\pi^\circ$,$\eta$, and $\pi\pi$ cross sections over a series of nuclei.
The studies have not
provided an obvious hint for a depletion of resonance structure.
The observed reduction and change of shape in the second resonance region
are mostly as expected from absorption effects, Fermi smearing and Pauli blocking,
and collisional broadening.
The solid line in Fig.~\ref{fig:1} is the sum of the available
meson cross sections between 400 and 800~MeV demonstrating the persistence
of the second resonance bump when at least one neutral meson is observed.
Here, it would be desirable to complete the picture by investigating single
charged pion as well as $\pi^+\pi^-$ production from nuclei.
In \cite{Krusche:2004zc}, it is suggested that there is
a large difference between quasifree meson
production from the nuclear surface and non-quasifree components.
The quasifree part scales with the nuclear surface and
does not show a suppression of the bump in the second
resonance region. Meanwhile, the (unobservable)
non-quasifree meson production would have larger contributions from the
nuclear volume.

However, in the reaction $\pi^\circ\pi^\pm$,
the two pions can stem from the decay of the $\rho$ meson while
the decay $\rho\to\pi^\circ\pi^\circ$ is forbidden.
Meanwhile, $\pi^\circ\pi^\circ$ pairs can stem from the decay
of the elusive $\sigma$ meson.
Accordingly, detailed studies of differential cross sections might reveal
different modifications of the $\pi\pi$ correlations.

\section{$\pi\pi$ Correlations in the Nuclear Medium}
\label{sec:pipiM}

The idea of strong threshold effects due to the $\pi \pi$ interaction in a
dense nuclear medium was first suggested in \cite{Schuck:1988jn}.
In the prediction of~\cite{Bernard:1987im},
a linear decrease with baryon density is assumed for moderate densities:
$m_\sigma = m_{\sigma_\circ} \cdot (1 - \alpha\cdot\rho/\rho_\circ)$.
A change in the shape of the invariant mass distribution is also
predicted for $\alpha=0$ as a result of the p-wave
coupling of pions to particle-hole and $\Delta$-hole states.
The in-medium behavior of scalar mesons is one of the key issues
for in-medium studies.
Here, the elusive $\sigma$ meson would be a prime candidate in the
search for a signature of chiral restoration
because it is the lightest meson possessing the same
quantum numbers as the QCD vacuum.

Some theoretical models expect a dropping of the $\sigma$ meson mass
as a function of nuclear density on account of partial restoration
of chiral symmetry
\cite{Lutz:1992dv,Hatsuda:1999kd,Rapp:1998fx}.
Recent theoretical papers consider this possibility
where the pion (J$^p=0^-$) and the $\sigma$ meson (J$^p=0^+$)
are regarded as chiral partners.
Several models describe the density dependence of the $\sigma$ and $\pi$ mass.
Being a Goldstone boson, the pion mass does not change dramatically
with density.
In order to reach the chiral limit of mass degeneracy,
the $\sigma$ mass would have to reduce.
With the main decay mode of the $\sigma$ meson being the decay into pion pairs,
a number of authors have performed calculations for the expected mass
distributions predicting sizeable $\pi\pi$ mass shifts already at
normal nuclear densities.

Roca, Oset et al. interpret the $\sigma$
meson as a scalar $\pi\pi$ scattering resonance and predict
a decrease of the $\pi\pi$ invariant mass with increasing nuclear density,
resulting from an in-medium modification of the $\pi\pi$
interaction~\cite{Roca:2002vd}.
Here, the meson-meson interaction in the scalar-isoscalar channel
is studied in the framework of  a chiral unitary approach
at finite baryon density.
The calculation dynamically generates the $f_0$ and $\sigma$ resonances
reproducing the meson-meson phase shifts in vacuum.
These theoretical results also
find a drop of the $\sigma$ resonance pole together
with a reduction of the resonance width in the nuclear medium.
In this case, the basic ingredient driving the mass decrease is
the p-wave interaction of the pion with the baryons in the medium.

In-medium modifications of the $\pi\pi$ interaction have been
studied in pion-induced reactions on nuclei like
A($\pi^+$,$\pi^+\pi^-$)~\cite{Bonutti:2000bv} and
A($\pi^-$,$\pi^\circ\pi^\circ$)~\cite{Starostin:2000cb}.
However, pion-induced reactions occur at
fractions of the normal nuclear density.
This complicates the interpretation of the data in terms of medium
effects.

Photon-induced reactions can reach normal nuclear densities
and should thus be more sensitive to in-medium modifications.
A first result came from the investigation of $\pi\pi$ invariant mass
distributions in the incident photon energy range of
400--460~MeV~\cite{Messchendorp:2002au}.
In this energy regime, the final state pions undergo some absorption and
little final state interactions, like rescattering.
The results indicate an effect consistent with a significant in-medium
modification in the $A(\gamma,\pi^\circ\pi^\circ)$ ($I$=$J$=0) channel.
\begin{figure}[htb]
  \epsfxsize=0.7\linewidth
\centerline{\epsfbox{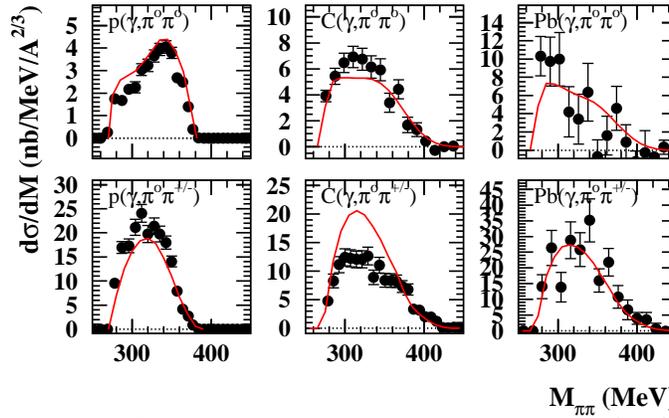}}
\caption{
 Differential cross sections of the reaction
 $A(\gamma,\pi^\circ\pi^\circ)$ (left) and
 $A(\gamma,\pi^\circ\pi^\pm)$ (right)
 with $A$=$^1$H,$^{12}$C,$^{\rm nat}$Pb
 for incident photons in the energy range of 400-460~MeV
  \protect\cite{Messchendorp:2002au}.
 Solid lines are the predictions from \protect\cite{Roca:2002vd}.
}\label{fig:sigma-exp}
\end{figure}
Figure~\ref{fig:sigma-exp} shows the threshold $\pi\pi$
production on nuclei measured with the TAPS spectrometer at MAMI-B.
The systematics includes p, C, Ca, and Pb.
They reveal a shape change of the $\pi^\circ\pi^\circ$ invariant mass
with increasing mass number as predicted in \cite{Roca:2002vd}
and would also be consistent with a dropping of the
$\sigma$ meson in medium.
It was confirmed that another isospin channel, here $\pi^\circ\pi^\pm$,
does not show such a behavior (lower row of Fig.~\ref{fig:sigma-exp}).
A rigorous comparison to theoretical predictions
could shed light on the nature of the $\sigma$ meson.

\section{Summary and Outlook}

The systematic study of the total production cross sections for single
$\pi^\circ$, $\eta$, and $\pi\pi$ cross sections over a series of nuclei has not
provided an obvious hint for a depletion of resonance yield.
The observed reduction and change of shape in the second resonance region
are mostly as expected from absorption effects, Fermi smearing and Pauli blocking,
and collisional broadening.
It has to be concluded that the medium modifications leading to the depletion of
cross section in nuclear photoabsorption are a subtle interplay of effects.
Their investigation and the rigorous comparison to theoretical models requires
a detailed study of differential cross sections and a deeper understanding of
meson production in the nuclear medium.

One such study investigates the possible change in the correlation
between low-momentum pion pairs in the nuclear environment.
First results have been presented and are found to be consistent with
a significant in-medium
modification in the $A(\gamma,\pi^\circ\pi^\circ)$ ($I$=$J$=0) channel.
For a rigorous comparison to theoretical predictions, improved statistics
on an extended systematics of nuclei is being acquired \cite{prop-pipi-nucs}.

\end{document}